\documentclass[aps,prb,twocolumn,showpacs,showkeys,floadfix,groupedaddress]{revtex4}
\usepackage{bm}
\usepackage{bbm}
\usepackage{amsmath}
\usepackage{amssymb}
\usepackage{t1enc}
\usepackage[cp1250]{inputenc}
\usepackage[english]{babel}
\usepackage{graphicx}

\begin{document}

\title{Transition from ballistic to diffusive behavior of graphene ribbons \\
in the presence of warping and charged impurities}
\author{J. W. K{ł}os$^{1,2}$}
\email{klos@amu.edu.pl}
\affiliation{$^{1}$Surface Physics Division, Faculty of Physics, Adam Mickiewicz
University,\\
Umultowska 85, 61-614 Poznań, Poland\\
$^{2}$Solid State Electronics, Department of Science and Technology, Linkö%
ping University 601 74, Norrköping, Sweden }
\author{A. A. Shylau$^{2}$}
\affiliation{$^{1}$Surface Physics Division, Faculty of Physics, Adam Mickiewicz
University,\\
Umultowska 85, 61-614 Poznań, Poland\\
$^{2}$Solid State Electronics, Department of Science and Technology, Linkö%
ping University 601 74, Norrköping, Sweden }
\author{I. V. Zozoulenko$^{2}$}
\email{Igor.Zozoulenko@itn.liu.se}
\affiliation{$^{1}$Surface Physics Division, Faculty of Physics, Adam Mickiewicz
University,\\
Umultowska 85, 61-614 Poznań, Poland\\
$^{2}$Solid State Electronics, Department of Science and Technology, Linkö%
ping University 601 74, Norrköping, Sweden }
\author{Hengyi Xu}
\affiliation{Condensed Matter Physics Laboratory, Heinrich-Heine-Universität, Universitä%
tsstrasse 1, 40225 Düsseldorf, Germany}
\author{T. Heinzel}
\affiliation{Condensed Matter Physics Laboratory, Heinrich-Heine-Universität, Universitä%
tsstrasse 1, 40225 Düsseldorf, Germany}

\begin{abstract}
We study the effects of the long-range disorder potential and warping on the
conductivity and mobility of graphene ribbons using the Landauer formalism
and the tight-binding \textit{p}-orbital Hamiltonian. We demonstrate that as
the length of the structure increases the system undergoes a transition from
the ballistic to the diffusive regime. This is reflected in the calculated
electron density dependencies of the conductivity and the mobility. In particular, we
show that the mobility of graphene ribbons varies as $\mu (n)\sim
n^{-\lambda },$ with $0\leq \lambda \lesssim 0.5.$ The exponent $\lambda $
depends on the length of the system with $\lambda =0.5$ corresponding to
short structures in the ballistic regime, whereas the diffusive regime $%
\lambda =0$ (when the mobility is independent on the electron density) is reached for
sufficiently long structures. Our results can be used for the interpretation
of experimental data when the value of $\lambda $ can be used to distinguish
the transport regime of the system (i.e. ballistic, quasi-ballistic or
diffusive). Based on our findings we discuss available experimental results.
\end{abstract}

\maketitle



\section{Introduction}

The two-dimensional allotrope of carbon - graphene has become a subject of
intensive research since its isolation in 2004.\cite{Novoselov} This is
because of its fundamental significance, its unusual electronic properties
as well as its potential for numerous applications (for a review see e.g.
\cite{review}).

A very interesting and not fully resolved problem is the impact of various
mechanisms such as disorder, substrate, environment, etc., on the transport
properties of graphene. Particular attention has been devoted to studies of
the effect of charged impurities (located in the substrate or on its surface) which is
widely considered to be the main mechanism limiting the mobility in graphene.%
\cite{Ando,Nomura2006,Adam,Hwang2007,Nomura2007} Indeed, recent experiments
on suspended graphene sheets have demonstrated a significant improvement of
electrical transport in suspended devices compared to traditional samples
where the graphene is supported by an insulating substrate.\cite{Bolotin,Du}

Another important mechanism that can affect the transport in graphene is
warping, when the graphene, as an elastic membrane, tends to become rippled
in order to minimize the elastic energy.\cite%
{Meyer,Fasolino,Kim_Castro_Neto,Katsnelson,Juan,Isacsson,Geringer,Deshpande}
The warped character of a graphene surface has been proved in diffraction
experiments\cite{Meyer} and STM measurements\cite{Geringer,Deshpande} for
both suspended samples and samples on an insulating substrate. It has been
demonstrated that due to the rehybridization effects and the change in the
next-to-nearest-neighbor hopping integrals caused by curvature the warping generates
spatially varying potential that is proportional to the square of the local
curvature.\cite{Kim_Castro_Neto}

The transport properties of the graphene can also be strongly affected by
its interaction with the substrate and other materials which may exist in
its environment.\cite{Sabio} This includes e.g. the interaction of the
graphene with the surface polar modes of SiO$_{2}$ or with water molecules
that might reside on the surface. These interactions have a long-range
character, and because of the corrugated character of the graphene and/or
dielectric surfaces, the spatial variation of these interactions would
result in a spatially varying effective potential affecting the transport
properties of the graphene sheet.

The purpose of the present paper is twofold. Our first aim is to study the
effect of warping on the transport properties of graphene ribbons. The
warping of the graphene affects both the next-neighbor and the
next-to-nearest-neighbor hopping integrals, $t$ and $t^{\prime }$
respectively. Previous works dealt primarily with the effect of the
modification of the next-to-nearest-neighbor hopping integrals.\cite%
{Kim_Castro_Neto,Isacsson} This is because the next-to-nearest-neighbor
hopping integrals $t^{\prime }$ are are much more strongly affected by
out-of-plane deformations in comparison to the nearest-neighbor integrals $%
t. $ On the other hand, because the electronic and transport properties of
graphene are primarily determined by the next-neighbor hopping, it is not
\textit{a priori} clear which effect is dominant. In this work we, based on
the realistic model of a warped graphene surface and the tight-binding $p$%
-orbital Hamiltonian, numerically study the effect of modification of the
next-neighbor hopping integrals $t$ on the conductance of the graphene ribbons. We
find that the modification of the nearest-neighbor hopping integrals due to the
out-of-plane deformations of the graphene surface has a negligible effect on
the conductance in comparison to the effect of charged impurities even for
moderate strength and concentration.

The second and the main aim of our study is the investigation of the
transition from the ballistic to diffusive behavior of graphene ribbons with
a realistic long-range disordered potential. One of the motivations for this
study are recent experiments addressing the mobility of suspended and
nonsuspended graphene devices of submicrometer dimensions. The dimension of
these devices is smaller than the phase coherence length $l_{\phi }$ at the
low temperature ($l_{\phi }\sim 3-5$ $\mu $m at 0.25 K and $\sim 1$ $\mu $m
at 1 K)\cite{Miao,Russo} and the mean free path approaches its ballistic
value\cite{Du}. This indicates that these submicrometer devices can be in
quasi-ballistic and even ballistic transport regime requiring the Landauer
approach for the description of the transport. At the same time, the
electron transport in these devices was analyzed in terms of the classical
mobility $\mu $ which is appropriate for a diffusive transport regime. In
the present study we use a realistic model of a disordered potential and the
tight-binding $p$-orbital Hamiltonian, and perform numerical calculations of
the conductance of graphene ribbons based on the Landauer formalism. We
demonstrate that as the size of the system $L$ increases the system
undergoes a transition from the ballistic to the diffusive regime. This is
reflected in the calculated electron density dependency of the conductivity and the
mobility. In particular, we show that the mobility of graphene ribbons
varies as $\mu (n)\sim n^{-\lambda },$ with $0\leq \lambda \lesssim 0.5.$
The exponent $\lambda $ depends on the size of the system with $\lambda =0.5$
corresponding to short structures in the purely ballistic regime, whereas
the diffusive regime corresponds to $\lambda =0$ (when the mobility is
independent on the electron density) and is reached for sufficiently long structures.
Our results can be used for the interpretation of experimental data when the
value of the parameter $\lambda $ can distinguish the transport regime of
the system (i.e. ballistic, quasi-ballistic or diffusive).

It should be noted that various aspects of the effect of the disorder on the
electron transport in graphene have been extensively studied in the past.%
\cite%
{Bardarson,Rycerz,Areshkin,Louis,Gunlycke,Li,Querlioz,Avouris,Martin,Lewenkopf,Mucciolo,Xu_bilayer,Rossi}
We stress that the focus of our study is the understanding of the transition
from the ballistic to the diffusive regime when the obtained electron density
dependencies for the conductivity and the mobility can be used to extract
information on the character of the transport regime of the system at hand.
Note that in contrast to many previous studies focussing on the
metal-insulator transition and the strong localization regime, in the
present paper we consider the case a ribbon with many propagating channels
when the localization length exceeds the size of the system. Note also that
because of computational limitations we keep in our study the width of the
ribbon $W$ constant and increase its length $L,$ such that the diffusive
regime is achieved when $L/W\gg 1$. We however expect that the results and
conclusions presented in this paper would remain valid even for bulk
diffusive samples with $L/W\sim 1.$

Finally, it is well established that edge disorder strongly affects the
transport properties of graphene ribbon.\cite%
{Bardarson,Rycerz,Areshkin,Louis,Gunlycke,Li,Querlioz,Avouris,Martin,Lewenkopf,Mucciolo,Xu_bilayer,Chen2007,Han2007}%
. However, the measurements of the mobility are typically done in the
multi-terminal Hall geometry where the edges do not play a role. Therefore,
in the present study we consider perfect edges to make sure that the
electron conductance is influenced only by the long-range potential in the
bulk. In our calculations we use the long-range potential corresponding to
remote charged impurities. We however demonstrate that the obtained results
are not particularly sensitive to the parameters of the potential. We
therefore can expect that our findings can be applicable not only to the
charged impurities, but to other mechanisms discussed above (e.g.
interaction with the surface polar modes, etc.) that can also be described
by a similar long-range potential.

The paper is organized as follows. In Sec. II A we present the basics of our
computational method for calculation of the conductance and the mobility of
graphene ribbons. The models of warping and remote impurities are described
in Sec. II B and Sec. II C. The conductivity and the mobility of graphene
ribbons in the presence of warping and charged impurities are presented and
discussed in Sec. III. Section IV contains the summary and conclusions.

\section{Model}

\begin{figure}[tbp]
\centering
\includegraphics[keepaspectratio, width = 0.8\columnwidth]{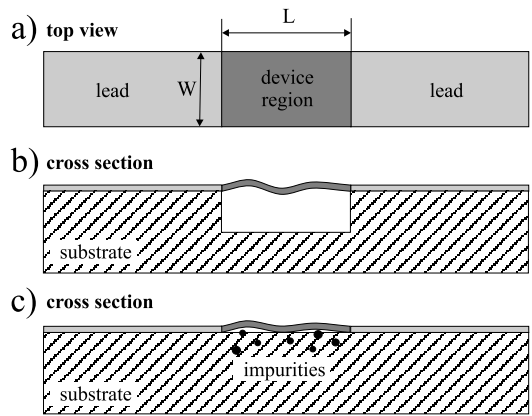}
\caption{The schematic sketch presenting the structure under consideration:
(a) top view of the device region (dark gray area) attached to the
semi-infinite graphene leads (light gray areas), (b) free standing, rippled
graphene layer, (c) graphene layer supported on the substrate (white area)
in the presence of charged impurities (dots). }
\label{fig:f1}
\end{figure}

\subsection{Basics}

In order to describe transport and electronic properties of graphene we use
the standard $p$-orbital tight-binding Hamiltonian,
\begin{equation}
H=\sum_{i}V_{i}\left\vert i\right\rangle \left\langle i\right\vert
-\sum_{i,j}t_{i,j}\left\vert i\right\rangle \left\langle j\right\vert
\end{equation}%
limited to the nearest-neighbor hopping. $V_{i}$ denotes the external
potential at the site $i$; the summation of $i$ runs over the entire
lattice, while $j$ is restricted to the sites next to $i$. We related the
spatial variation of the hopping integral $t_{i,j}$ with bending and
stretching of the graphene layer due to warping as described in the next
section.

The conductance $G$ and the electron density $n$ are computed with the
aid of recursive Green function technique \cite%
{Zozoulenko1996,Zozoulenko2008}. We assume that the semi-infinite leads are
perfect graphene ribbons, and the device region is a rectangular graphene
strip and the imperfection (warping and long-range impurity potential) are
restricted only to this area (see Fig.\ref{fig:f1}). The zero-temperature
conductance $G$ is given by the Landauer formula
\begin{equation}
G=\frac{2e^{2}}{h}T,
\end{equation}%
where $T$ is the total transmission coefficient between the leads. Then we
calculated the conductivity
\begin{equation}
\sigma =\frac{L}{W}G,  \label{eq:e01}
\end{equation}%
the electron density
\begin{equation}
n(E)=\int_{0}^{E}dE\;\mathrm{DOS}(E),  \label{eq:e02}
\end{equation}%
and the mobility
\begin{equation}
\mu =\frac{\sigma }{en},  \label{eq:e03}
\end{equation}%
as a functions of the Fermi energy $E$ ($W$ and $L$ denote width and length
of device, respectively). The density of states (DOS) was computed by
averaging the local density of states (LDOS) over the whole device area. The
LDOS is given by the diagonal elements of the total Green's function.\cite%
{Zozoulenko2008}

All the results presented here correspond to the ribbons of the zigzag
orientation. Previous studies do not show a difference of the transport
properties of the zigzag and armchair ribbons in the presence of disorder
(provided the disorder concentration is sufficiently high).\cite%
{Martin,Xu_bilayer} We therefore expect that all the results reported here
remain valid for the case of the armchair orientation as well.

The effect of warping is included in our model by modification of hopping
integrals $t$ resulting from stretching (contraction) and $\pi -\sigma $
rehybridisation. The external potential of remote impurities (with inclusion
the effect of screening by carries) is reflected in the model by changing of
site energies. In the next subsection we present a detailed description of
models for the warping and remote impurities.

\subsection{Corrugation}

\begin{figure}[tbp]
\centering
\includegraphics[keepaspectratio, width = \columnwidth]{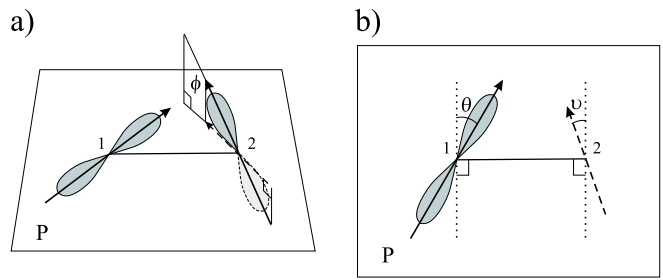}
\caption{Definition of angles (used in Eq.(\protect\ref{eq:e1})) describing
alignment of $p$-orbitals of neighboring atoms. (a) $\protect\phi$ denotes
the angle between direction of orbital 2 and its projection (dashed arrow)
on the plane $P$ spanned on the direction of the orbital 1 and the vector
linking the centers of the orbitals 1 and 2; (b) Top view of the plane $P$.
All vectors depicted here belong to the plane $P$ and have the same meaning
as in (a). Dotted lines are perpendicular to the vector linking centers of
the orbitals 1 and 2. Note that in the panel (b) the orbital 2 is not shown.}
\label{fig:f2}
\end{figure}

The mechanical properties of graphene can be modelled by treating this
system as an elastic membrane. One can distinguish two modes of deformation:
stretching/contraction and bending. Both of them affect the strength of
carbon-carbon (C-C) bonds within the graphene sheet by changing the distance
between carbon atoms and the alignment of their $p$-orbitals, respectively.
In the tight-binding model the hopping integral corresponding the hybridised
$\sigma -\pi $ bond is given by,\cite{Hansson-Stafstrom,Mulliken}
\begin{eqnarray}
t(a,\theta ,\phi ,\vartheta ) &=&\cos (\vartheta )\cos (\theta )\cos (\phi
)t_{2p\pi ,2p\pi }(a)  \notag \\
&&-\sin (\vartheta )\sin (\theta )t_{2p\sigma ,2p\sigma }(a),  \notag \\
t_{2p\pi ,2p\pi }(a) &=&\alpha _{\pi }e^{a\zeta }(1+a\zeta +\tfrac{2}{5}%
(r\zeta )^{2}+\tfrac{1}{15}(r\zeta )^{3}),  \notag \\
t_{2p\sigma ,2p\sigma }(a) &=&\alpha _{\sigma }e^{a\zeta }(-1-a\zeta -\tfrac{%
1}{5}(a\zeta )^{2}+\tfrac{2}{15}(a\zeta )^{3}),  \label{eq:e1}
\end{eqnarray}%
where $\alpha _{\pi }/t_{0}\approx -4.23$, $\alpha _{\sigma }/t_{0}\approx
-4.33$ and $\zeta \approx 3.07$\AA $^{-1}$ ($t_{0}=t(a_{0},0,0,0)=2.7$ eV
and $a_{0}=1.42$ \AA\ are the hopping integral and the C-C bond length for
flat and unstained graphene ribbon, respectively). $t_{2p\pi ,2p\pi }$ and $%
t_{2p\sigma ,2p\sigma }$ denote the hopping integrals for pure $\pi $-bonds
(in flat graphene) and $\sigma $-bonds (for collinear p-orbitals). The
spatial orientation of the $p$-orbitals is described by the angles $\theta
,\phi ,\vartheta $ as presented in Fig.\ref{fig:f2}.

The analysis of diffraction patterns of corrugated graphene\cite{Meyer}
provides information about the range in which the normal to the surface
varies. The measured range $\pm 5^{o}$ allows to estimate from Eq.(\ref%
{eq:e1}) an impact of bending on the relative change of the hopping
integral, $t(a_{0},\theta ,\phi ,\vartheta )-t_{0})/t_{0}\approx 0.4\%$. The
information about the distribution of bond lengths in corrugated graphene is
provided by the Monte Carlo simulations.\cite{Fasolino} Using this data we
can calculate the change of hopping integrals (\ref{eq:e1}). The relative
change $(t(a,0,0,0)-t_{0})/t_{0}$ for C-C bond variation $\Delta a=a-a_{0}$,
corresponding to the half width at half height (HWHH) of the bond length
distribution \cite{Fasolino}, is approximately $2\%$. This comparison shows
that the effect of strain (leading to a change of the bond length $\Delta a$%
) has a stronger impact on the modification of the hopping integrals in
comparison to the effect of bending (related to the orbital alignment).

\begin{figure}[tbp]
\centering
\includegraphics[keepaspectratio, width = \columnwidth]{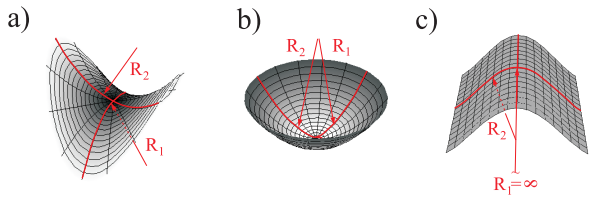}
\caption{(color online) (a) A saddle-like surface and (b) a local minimum
created by a nonuniform contraction (dark areas) and inflation (light areas)
of a flat region. (c) developable surface (i.e. the surface with zero Gauss
curvature). Each red curve and the vectors showing the radius of curvature
lie in the same normal plane.}
\label{fig:f3}
\end{figure}

In a corrugated free standing membrane the bending and the in-plain strain
are related to each other.\cite{Landau,Nelson,Witten} For example, in order
to produce a local minimum/maximum or a saddle-like area in a flat membrane
it is necessary to introduce a strain (see Fig.\ref{fig:f3}). Bending and
strain of the free standing membrane can be related to the Gauss curvature:
\begin{equation}
c_{G}(x,y)=c_{1}(x,y)c_{2}(x,y),
\end{equation}%
which is the product of the minimum $c_{1}$ and maximum $c_{2}$ of normal
curvatures\cite{Oprea} (called by principal curvatures). The curvatures $%
c_{1}$, $c_{2}$ are given by the inverse radiuses of the local curvatures, $%
|c_{1}|=1/R_{1}$, $|c_{2}|=1/R_{2}$, see Fig.\ref{fig:f3}. The sign of $%
c_{1} $ and $c_{2}$ depends on whether an intersection of normal plane with
the surface is convex or concave. The positive value of $c_{G}$ at some
point corresponds to the presence of a local minimum/maximun which requires
streching of this region. For a saddle-like area (and negative value of $%
c_{G}$) the local contraction of a flat membrane is needed. The only kind of
bending which does not produce a strain corresponds to developable surfaces
(i.e. surfaces with zero Gauss curvature) such as e.g. a cylindrical tube,
see Fig.\ref{fig:f3} (a).

Because of a small variation of the relative change of the bond length $%
\Delta a/a_{0}$ we assume a linear dependence of the bond
elongation/contraction $\Delta a$ on the Gauss curvature:
\begin{equation}
\Delta a(x,y)=kc_{G}(x,y).  \label{eq:e10}
\end{equation}%
The value of the coefficient $k$ was chosen to reproduce the range of $%
\Delta a$ variation computed in \cite{Fasolino}. We estimated the HWHH of $%
\Delta a$ from the the C-C bond length distribution presented in \cite%
{Fasolino}. Then, we computed the HWHH of $c_{G}$ distribution for a large
(generated as described below) corrugated ribbon. The coefficient $k$ was
defined as a ratio of the HWHHs for the $\Delta a$ and $c_{G}$ distributions.

In our model we do not take into account an influence of the surface
corrugation on the position of carbon atoms. The positions of the centres of
C-C bonds in flat lattice were projected onto a corrugated continuous
surface. In these points the Gauss curvature was computed. The value of $%
c_{G}$ was used to calculate the change of the bond lengths (\ref{eq:e10}).
The angles $\theta ,\phi ,\vartheta $ (see Fig.\ref{fig:f2}) were calculated
from the alignment of normals representing the $\pi $ orbitals. The normals
were computed at the ends of the projected bonds. The computed angles $%
\theta ,\phi ,\vartheta $ and C-C bond lengths $a=a_{0}+\Delta a$ were used
to calculate the next-neighbor hopping integrals (\ref{eq:e1}) in the
corrugated ribbon.

In order to model the geometry of a corrugated ribbon we used the
description of fluctuations of elastic membranes presented in\cite{Fasolino}%
. The ripples on the graphene surface can be expanded as a Fourier series of
plane waves characterized by the wave vectors ${\bm q}$. In the harmonic
approximation\cite{Fasolino} the in-plane- and out-of-plane displacement $h$
are decoupled. In this approximation the mean-square amplitude of Fourier
component $h_{q}$ is given by\cite{Nelson,Katsnelson}:
\begin{equation}
\left\langle h_{q}^{2}\right\rangle \sim \frac{1}{q^{4}}.  \label{eq:e11}
\end{equation}

\begin{figure}[tbp]
\centering
\includegraphics[keepaspectratio, width = 0.9\columnwidth]{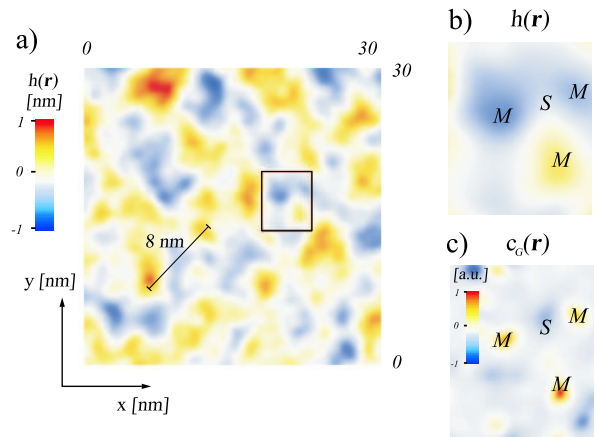}
\caption{(color online) The model of ripples on the graphene surface. The
black lines denote characteristic length of ripples 8nm ; (b) the magnified
small area marked in (a) by a black frame; M and S stand for local maxima
(minima) and saddle points, respectively; (c) the Gauss curvature
corresponding to a part of the surface presented in (b) - red/orange (blue)
areas mark regions with positive (negative) Gauss curvature. }
\label{fig:f4}
\end{figure}
It can be shown that the mean-square of the out-of-plane displacement $h$
scales quadratically with the linear size of the sample\cite{Fasolino}:
\begin{equation}
\left\langle h^{2}\right\rangle =\sum_{q}\left\langle h_{q}^{2}\right\rangle
\sim L^{2}.  \label{eq:crumpling}
\end{equation}%
According to Eq.(\ref{eq:crumpling}) the height of the ripples increases
quadratically with the sample size. It means that a large sample should be
crumpled ($\left\langle h^{2}\right\rangle \gtrsim L$) which contradicts the
experiments. The results consistent with the experiments where large sample
remains approximately flat, are reproduced by the Monte Carlo simulations%
\cite{Fasolino} where the dependence $\left\langle h_{q}^{2}\right\rangle
\sim q^{-4}$ (Eq. (\ref{eq:e11})) remains valid for short wavelengths only,
and saturates for the long wavelengths $\lambda =2\pi /q\gtrsim \lambda
^{\ast }\approx 8$ nm. This mechanism is responsible for the existence of
ripples of characteristic size $\sim \lambda ^{\ast }$, see Fig. \ref{fig:f4}%
.

Using this guidance we modeled the corrugated surface in the following way.
The surface $h({\bm r})$ was generated by a superposition of plane waves,
\begin{equation}
h({\bm r})=C\sum_{i}C_{q_{i}}\sin ({\bm q}_{i}\cdot {\bm r}+\delta _{i}),
\label{eq:e13}
\end{equation}%
where ${\bm r}=(x_{1},x_{2})$ is the in-plane position vector. The
directions $\varphi _{i}$ of the wave vectors ${\bm q}_{i}=q_{i}(\cos
(\varphi _{i}),\sin (\varphi _{i}))$ and the phases $\delta _{i}$ were
chosen randomly. The length of wave vectors $q_{i}$ covers equidistantly the
range $2\pi /L<q_{i}<2\pi /(3a_{0})$, where $L$ is a leading linear size of the
rectangular area and $a_{0}$ denotes the C-C bond length (we assume that $L\gg
\lambda ^{\ast }$). The amplitude of the mode was given by the harmonic
approximation (\ref{eq:e11}) $C_{q}=\sqrt{2\left\langle
h_{q}^{2}\right\rangle }$ for the wave length $\lambda <\lambda ^{\ast },$
otherwise it was kept constant and equal to $C_{q^{\ast }}$, where $q^{\ast
}=2\pi /\lambda ^{\ast }$. We introduced the normalisation constant $C$ to
keep the averaged amplitude of the out-of-plane displacement $\bar{h}=\sqrt{%
2\left\langle h^{2}\right\rangle }$ equal to the experimental values $\bar{h}%
\approx 1$ nm for typical sizes of samples.

In Fig. \ref{fig:f4} the graphene surface, generated by using the procedure
described above, is shown. Long-wave ripples of the size $\sim \lambda
^{\ast }\approx 8$ nm discussed above are clearly seen. A small area marked
in a box in Fig. \ref{fig:f4}a is enlarged in Fig. \ref{fig:f4}b. In this
region two minima (blue spots), one maximum (yellow spot) and one saddle
region (white area between color spots) are shown. The relation between the
geometry of the surface and its Gauss curvature can be visible in Fig. \ref%
{fig:f4}c. The regions of stretching (red and yellow spots) and contraction
(blue spot) in Fig.\ref{fig:f4}c correspond to the position of
minima/maximum ($M$) and saddle area ($S$) in Fig.\ref{fig:f4}b,
respectively.

\subsection{Long-range potential}

The simplest model of charged impurities corresponds to the short-range $%
\delta $-function scattering centres,
\begin{equation}
V_{i}=U_{\delta }\sum_{i^{\prime }=1}^{N_{imp}}\delta \left( {\bm r}_{i}-{%
\bm r}_{i^{\prime }}\right) ,  \label{eq:e14}
\end{equation}%
where $U_{\delta }$ is the strength of the individual scatterer, and the
summation over $i^{\prime }$ runs over all impurities $N_{imp}$ in the
device region. According to the Boltzmann transport theory the conductivity
is $\sigma =(e^{2}/h)(2E\tau /\hbar )$, where $\tau $ stands for the
scattering time. For the short range potential (\ref{eq:e14}) the scattering
time $\tau \sim 1/(E_{F}n_{imp}U_{\delta }^{2})$ (where $n_{imp}$ is the
impurity concentration).\cite{ando} In this model the calculated
conductivity is independent on the electron density $n,$ whereas the
mobility $\mu \sim 1/n$,\cite{Nomura2007} which contradicts the experimental
observations.\cite{Bolotin} This indicates that such a simplified model
might not be appropriate for a description of a realistic system.

A long-range character of the electrostatic interaction is included in the
model of bare Coulomb-like scattering centers,
\begin{equation}
V_{i}=\frac{1}{4\pi \epsilon _{r}\epsilon _{0}}\sum_{i^{\prime }=1}^{N_{imp}}%
\frac{e^{2}}{\left\vert {\bm r}_{i}-{\bm r}_{i^{\prime }}\right\vert },
\end{equation}%
where $\epsilon _{0}$ and $\epsilon _{r}$ stand respectively for the vacuum
and relative permittivities. For this potential the conductivity is
proportional to electron density, $\sigma \sim n/n_{imp}.$\cite{Nomura2007}.
However, the application of the bare Coulomb potential can be justified only
for low values of the electron density when the screening effects
limiting the range of the potential are negligible.

The simplest screened potential is given by the Thomas-Fermi approximation,%
\cite{Adam2008}
\begin{equation}
V_{i}=U_{TF}\sum_{i^{\prime }=1}^{N_{imp}}\frac{\exp \left( -\xi _{TF}|{\bm r%
}-{\bm r}_{i}|\right) }{\left\vert {\bm r}_{i}-{\bm r}_{i^{\prime
}}\right\vert },  \label{TF}
\end{equation}%
where the parameters $U_{TF}$ and $\xi _{TF}$ describe the strength and the
range of the scattering centres for the Thomas-Fermi potential. The
inclusion of screening allows to achieve both the limits of Coulomb
scattering (for low $n$) and short-range scattering (for high $n$).

The singularity at ${\bm r}_{i}={\bm r}_{i^{\prime }}$ in the Thomas-Fermi
potential (\ref{TF}) can cause numerical difficulties. In our calculation we
utilize a model for screened scattering centers of the Gaussian shape
commonly used in the literature where the potential on the site $i$ reads%
\cite{Bardarson,Rycerz,Lewenkopf},
\begin{equation}
V_{i}=\sum_{i^{\prime }=1}^{N_{imp}}U_{i^{\prime }}\exp \left( -\frac{|{\bm %
r_{i}}-{\bm r}_{i^{\prime }}|^{2}}{2\xi ^{2}}\right) ,  \label{Gauss}
\end{equation}%
where the height of scattering centers is uniformly distributed in the range
\begin{figure}[tbp]
\centering
\includegraphics[keepaspectratio, width =
0.65\columnwidth]{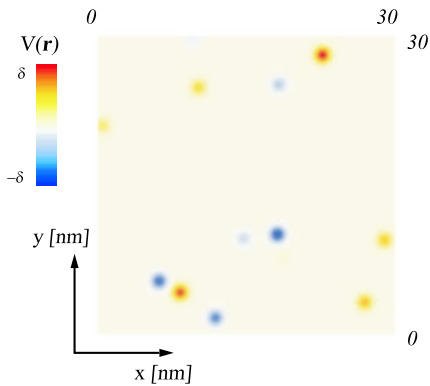}
\caption{(color online) Illustration of the long-range potential, Eq.(%
\protect\ref{Gauss}). The screening range is $\protect\xi =4a_{0}$, the
impurity concentration is $n_{imp}=10^{12}\,$cm$^{-2}$.}
\label{fig:f5}
\end{figure}
$U_{i^{\prime }}\in \lbrack -\delta ,\delta ]$. The strength and correlation
between the scattering centers is described by the dimensionless correlator $%
K$,\cite{Rycerz,Lewenkopf}
\begin{equation}
\left\langle V_{i}V_{j}\right\rangle =\frac{K(\hbar v_{F})^{2}}{2\pi \xi ^{2}%
}\exp \left( -\frac{|{\bm r_{i}}-{\bm r}_{j}|^{2}}{2\xi ^{2}}\right) ,
\end{equation}%
(note that $\left\langle V_{i}\right\rangle =0$). The averaging is preformed
for all possible configurations of system which differ only in the
distribution of the position of the impurities ${\bm r}_{i^{\prime }}$ and
the strength $U_{i^{\prime }}$ ($\delta $, $\xi $, $n_{imp}$ are kept
constant).

For this impurity potential both the screening range $\xi $ and the impurity
strength $\delta $ are independent on the electron density and play 
the role of parameters. They are related to $K$ by following formula\cite%
{Rycerz}:
\begin{equation}
K\approx 40.5\tilde{n}_{imp}(\delta /t)^{2}(\xi /a_{0}),  \label{eq:e20}
\end{equation}%
where $\tilde{n}_{imp}=N_{imp}/N$ denotes the relative concentration
expressed via the total number of impurities $N_{imp}$ and the atomic sites $%
N$ in a sample. In our calculation we use $n_{imp}=10^{12}$ cm$^{-2}$ as a
typical value given in the experiment\cite{Du}. We assumed that the
screening range\ $\xi =4a_{0}$. The strength of impurities was chosen in
order to get typical values of correlator\cite{Hwang2007,Bardarson} (we
chose $K=1,2,4,8$).

\section{Results and discussion}

\begin{figure}[tbp]
\centering
\includegraphics[keepaspectratio, width = 0.9\columnwidth]{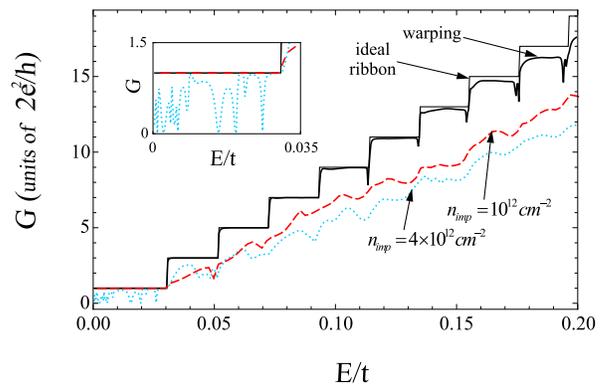}
\caption{(color online) Effect of warping and charged impurities on the
conductance of graphene nanoribbon of length $L=30.6$ nm and width $W=31.8$
nm ($250\times 150$ sites). The thin line shows the conductance steps for an
ideal ribbon. The bold line denotes the conductance for a system with
warping only; dashed and dotted lines refer to the conductance of a ribbon
with impurities (without warping) with the concentration $n_{imp}=10^{12}\,$%
cm$^{-2}$ and $4\times 10^{12}$cm$^{-2}$ respectively. The strength of the
impurities is $\protect\xi=4a$, $K=2$. The inset shows an enhanced view of
the conductance in the vicinity of the first conductance step.}
\label{fig:warping}
\end{figure}

Let us start with the comparison of the effects of warping and charged
impurities on the conductance of a graphene ribbon. Figure \ref{fig:warping}
shows the calculated conductance of a representative ribbon of the dimension
$\sim 31\times 32$ nm. The warping modifies the conductance only slightly.
For low energies close to the charge neutrality point the conductance steps
remain practically unaffected. For higher energies the conductance steps
become somehow distorted with the conductance plateaus being gradually
shifted down and sharp minima appearing next to the rising edges of the
plateaus. At the same time, in the presence of charged impurities even of a
moderate strength the conductance of the ribbon is distorted substantially.
The conductance steps are significantly washed out and the overall slope of
the curve is lowered. Note that Fig. \ref{fig:warping} shows the conductance
of the ribbon for two representative impurity configurations with $%
n_{imp}=10^{12}$ and $4\times 10^{12}$ cm$^{-2}$ (dashed and dotted lines
respectively). For higher energies (outside the first conductance step) the
conductance for both cases shows rather similar features. However, for the
energies corresponding to the the first conductance plateau the conductance
of the sample with $n_{imp}=10^{12}$ cm$^{-2}$ (in contrast to the case of $%
n_{imp}=$ $4\times 10^{12}$ cm$^{-2})$ remains practically unaffected. This
is due to the fact that in the zigzag ribbons the transport in the vicinity
of the charge neutrality point takes place via edge states strongly
localized near the ribbon's boundaries. For the considered impurity
configuration for $n_{imp}=10^{12}$ cm$^{-2}$ no individual impurities are
located close to the boundaries, and, as a result, the first conductance
step remains practically unaffected.

Our results demonstrate that the modification of the nearest-neighbor
hopping integrals $t$ resulting from stretching (contraction) and $\pi
-\sigma $ rehybridisation has a little effect on the conductance in
comparison to the effect of impurities. It should be also noted that the
effect of the modification of the next-to-nearest hopping integrals $%
t^{\prime }$ is also weak in comparison to a realistic impurity potential.
For example, Kim and Castro Neto\cite{Kim_Castro_Neto} estimated that the
variation of the effective electrochemical potential generated by a spatial
variation of $t^{\prime }$ is of the order of 30 meV, which is an order of
magnitude smaller than a corresponding variation of the impurity potential%
\cite{Lewenkopf,Ando}. Hence, in our further analysis of the conductivity
and the mobility of graphene ribbons we will take into to account the impact
of charged impurities only.

\begin{figure}[tbp]
\centering
\includegraphics[keepaspectratio,
width=0.9\columnwidth]{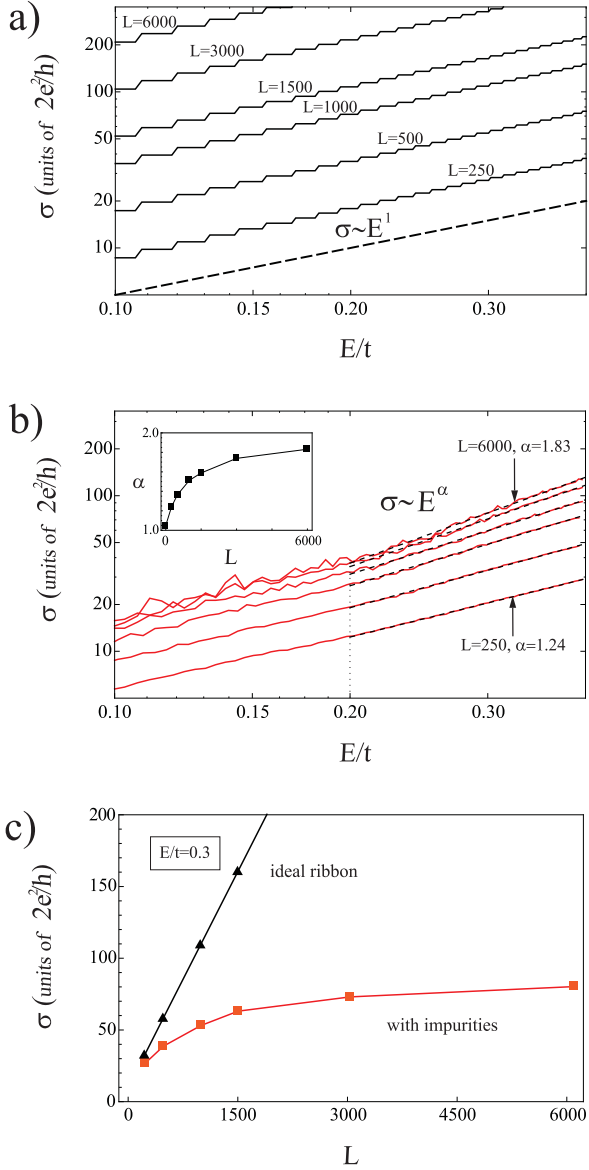} \caption{(color online) (a) The
conductivity of (a) ideal ballistic nanoribbons and (b) nanoribbons with impurities as a
function of the Fermi energy $E$. The nanoribbon lengths are $L=31,61,123,184,369,738$ nm
corresponding to 250,500,1000,1500,3000,6000 sites. The conductivities in (b) are
averaged over 10 impurities configuration. The dashed lines show the fit $\protect\sigma
\sim E^{\protect\alpha}$ for the energies $E>0.2t$ as
indicated by a vertical dotted line. (Here and hereafter we choose $E>0.2t$ because for lower
energies the fitted dependencies deviate from the power-low behavior due to sample-specific
fluctuations). The inset shows a dependence $\protect%
\alpha=\protect\alpha(L)$. The impurity parameters are $%
n_{imp}=10^{12}cm^{-2}$, $K=2$, $\protect\xi =4a$. (c) The conductivity as a
function of the ribbon length $L$ for ballistic ribbons and ribbons with
impurities. The ribbon width is $W=53.1$ nm (250 sites). }
\label{fig:conductivity}
\end{figure}

\begin{figure}[tbp]
\centering
\includegraphics[keepaspectratio, width = 0.9\columnwidth]{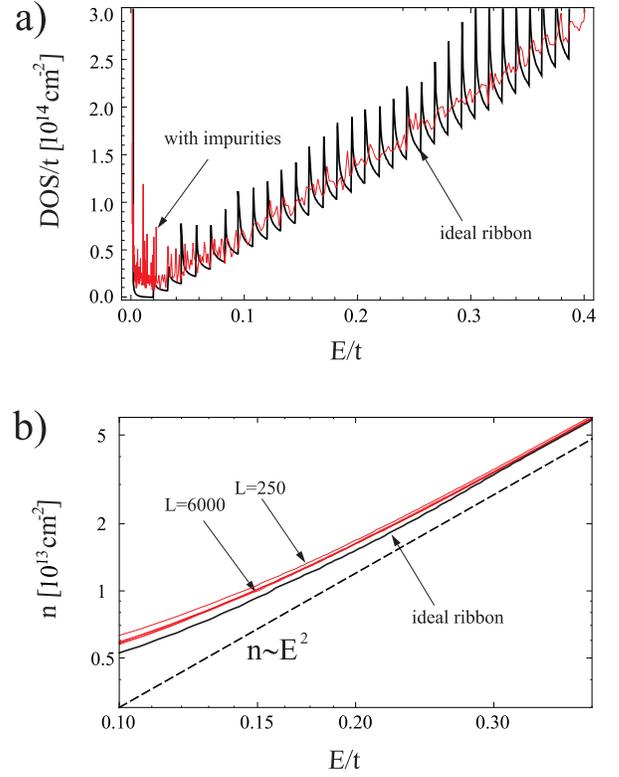}
\caption{(color online) (a) The DOS as a function of the Fermi energy of an
ideal ribbon (solid line) and of a ribbon of the length $L=184$ nm (1500
sites) for a representative impurity configuration. (b) The dependence of
the electron density $n$ on the Fermi energy for an ideal ribbon and for
ribbons of various length with impurities. The parameters of the ribbons and
the impurity strength are the same as in Fig. \protect\ref{fig:conductivity}. (Note that the electron densities for the ribbons with the length $L>250$ sites are
almost indistinguishable).}
\label{fig:DOS}
\end{figure}

\begin{figure}[!]
\centering
\includegraphics[keepaspectratio, width =
0.9\columnwidth]{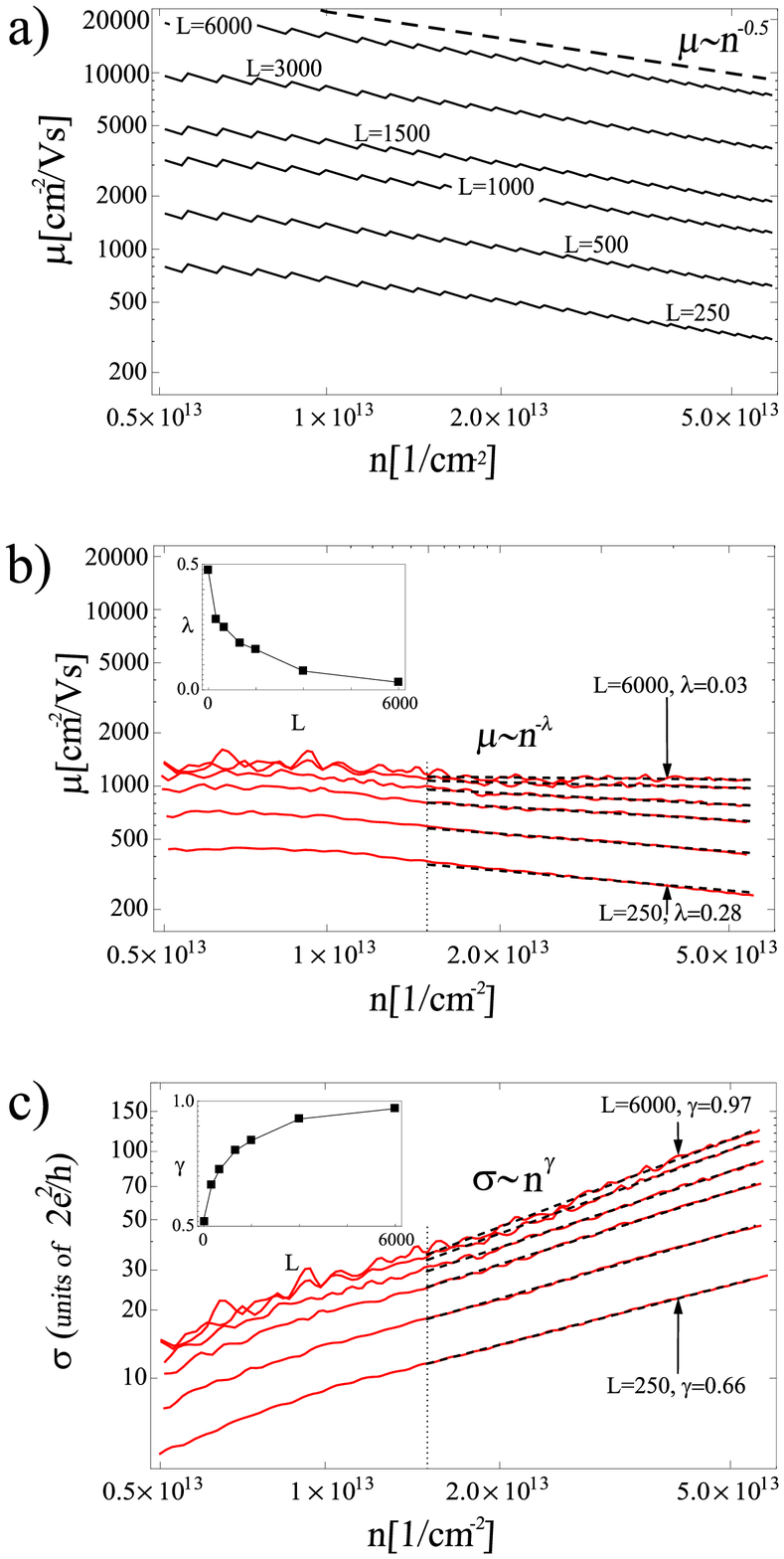}
\caption{(color online) The mobility as a function of the electron
density for (a) prefect zigzag nanoribbons and for (b) zigzag
nanoribbons ribbon in the presence of charged impurities. The inset in (b)
shows a dependence $\protect\lambda=\protect\lambda(L)$. (c) The
conductivity a function of the electron concentration for nanoribbons in the
presence of charged impurities. The inset in (c) shows a dependence $\protect%
\gamma=\protect\gamma(L)$. The parameters of the ribbons and the impurity
strength are the same as in Figs. \protect\ref{fig:conductivity}, \protect
\ref{fig:DOS}. The dashed lines in (b) and (c) show the fits $\protect\mu %
\sim \protect\mu^{\protect\lambda}$ and $\protect\sigma \sim \protect\sigma^{%
\protect\gamma}$ for the electron densities $n\gtrsim 1.5\times 10^{13}$ cm$^{-2}$
(as indicated by a vertical dotted line). The mobilities and conductivities
are averaged over 10 impurity configurations.}
\label{fig:mobility}
\end{figure}

Let us now turn to the investigation of transition from the ballistic to
diffusive regime of transport taking place as the size of the system
increases. Because of computational limitations we keep in our study the
width of the ribbon $W$ constant and increase its length $L,$ such that the
diffusive regime is achieved when $L/W\gg 1$. We however expect that the
results and conclusions presented in this paper would remain valid even for
bulk diffusive samples with $L/W\sim 1.$

Figure \ref{fig:conductivity} shows the conductivity $\sigma $ of ideal
ballistic graphene ribbons of different lengths $L$ exhibiting a linear
dependence on the electron energy $E$. This behavior of the conductivity
reflects the corresponding linear dependence of the conductance $G$ of
zigzag ribbons as a function of energy (see Appendix for details). Note that
the conductance of an ideal ribbon in the ballistic regime is independent on
the length of the system. Therefore, for fixed energy, the conductivity $%
\sigma =\frac{L}{W}G$ increases linearly with the ribbon length $L$ as
illustrated in Fig. \ref{fig:conductivity} (c). Apparently, in the ballistic
regime, the conductivity does not exist as a local property and can not be
considered as material parameter because it is size-dependent.

The conductivity of the ribbons of different lengths as a function of the
electron energy in the presence of impurities is shown in Fig. \ref%
{fig:conductivity} (b). The conductivity is strongly reduced in comparison
to the ideal conductance steps and is no longer a linear function of the
electron energy. Instead, it follows a power-law dependence
\begin{equation}
\sigma \sim E^{\alpha },  \label{scaling_sigma}
\end{equation}%
where the exponent $\alpha $ approaches $\alpha =2$ for sufficiently long
ribbons (see inset to Fig. \ref{fig:conductivity} (b)). Figure \ref%
{fig:conductivity} (c) shows the dependence of the conductivity $\sigma $ on
the length of the system in the presence of impurities. This dependence
exhibits a clear saturation of the conductivity for sufficiently large
systems, $L\gtrsim 6000$ sites (740 nm). In order to understand these
features, in particular, the dependence of the conductivity on the system
size $L,$ let us recall that the conductance of a disordered system is
expected to obey the scaling law
\begin{equation}
\ln (1+1/g)=L/\xi _{loc},  \label{scaling}
\end{equation}%
where $\xi _{loc}$ is the localization length and $g=G/G_{0}$ is the
dimensionless conductance ($G_{0}=2e^{2}/h$ being the conductance unit) \cite%
{Anderson}. In this study we focus on the transport regime with many
transmitted channels $g\gg 1.$ It follows from Eq. (\ref{scaling}) that this
corresponds to the case $\xi _{loc}\gg L,$ i.e. the localization length
exceeds the size of the system and the conductance is inversely proportional
to the length, $G\sim G_{0}/L.$ In this transport regime referred to as a
diffusive (or Ohmic), the conductivity $\sigma $ is therefore independent on
the system size and can be regarded as a local quantity. According to Fig. %
\ref{fig:conductivity} (c) this transport regime is achieved for
sufficiently long ribbons, $L\gtrsim 6000$ sites (740 nm). For smaller $L$
the system is in a quasi-ballistic regime when the conductivity $\sigma $
depends on the size of the system. In this case the conductance $G$ is
apparently more appropriate quantity to describe the transport properties of
the system at hand.

Let us now turn to the analysis of the mobility of graphene ribbons $\mu $.
For a classical (Ohmic) conductor the mobility is the fundamental material
property independent on the system size and the electron density. In
contrast, for ballistic ribbons the mobility is size-dependent and decreases
with increase of the electron density as $\mu \sim n^{-0.5}$ (see Appendix).
Therefore one can expect that the mobility of ribbons varies as
\begin{equation}
\mu (n)\sim n^{-\lambda },  \label{scale_mobility}
\end{equation}%
where the exponent $\lambda $ ranges from $0$ in the diffusive limit to $0.5$
in the ballistic limit.

In order to calculate the electron mobility $\mu =\sigma /en$ we, in
addition to the conductivity $\sigma $, have to calculate the electron
density $n$ in the ribbons (note that this step represents the most
time-consuming part of our numerical calculations because in order to
calculate $n$ we have to compute the DOS for all energies $0<E<E_{F}$).
Figure \ref{fig:DOS} shows the DOS for a representative impurity
configuration and electron densities for ribbons of different lengths
calculated from the DOS according to Eq. (\ref{eq:e02}). For ideal ribbons,
the DOS follows an overall linear dependence on the energy with the
singularities corresponding to openings of new propagating channels
characteristic for quasi-1D systems. Because of this linear increase of the
DOS, the electron density for an ideal ribbon follows a quadratic dependence on the
energy, $n\sim E^{2}$ (see Appendix). Figure \ref{fig:DOS} (a) shows that
the impurities only smear out the singularities in the DOS of an ideal
ribbon, but do not reduce the DOS. Therefore, regardless of the ribbon
lengths the average electron density in the ribbons with impurities is not reduced in
comparison to the ideal ribbons, and follows the same quadratic dependence $%
n\sim E^{2}.$ This behavior is expected for the transport regime at hand
when the localization length exceeds the size of the system, $\xi _{loc}\gg
L,$ such that the averaged local DOS is essentially independent of the size
of the system. Note that in the opposite regime of the strong localization, $%
\xi _{loc}\ll L,$ the electron density in ribbons is strongly reduced due to the
effect of impurities.\cite{Xu_bilayer}.

\begin{figure}[tpb]
\centering
\includegraphics[keepaspectratio, width = 0.9\columnwidth]{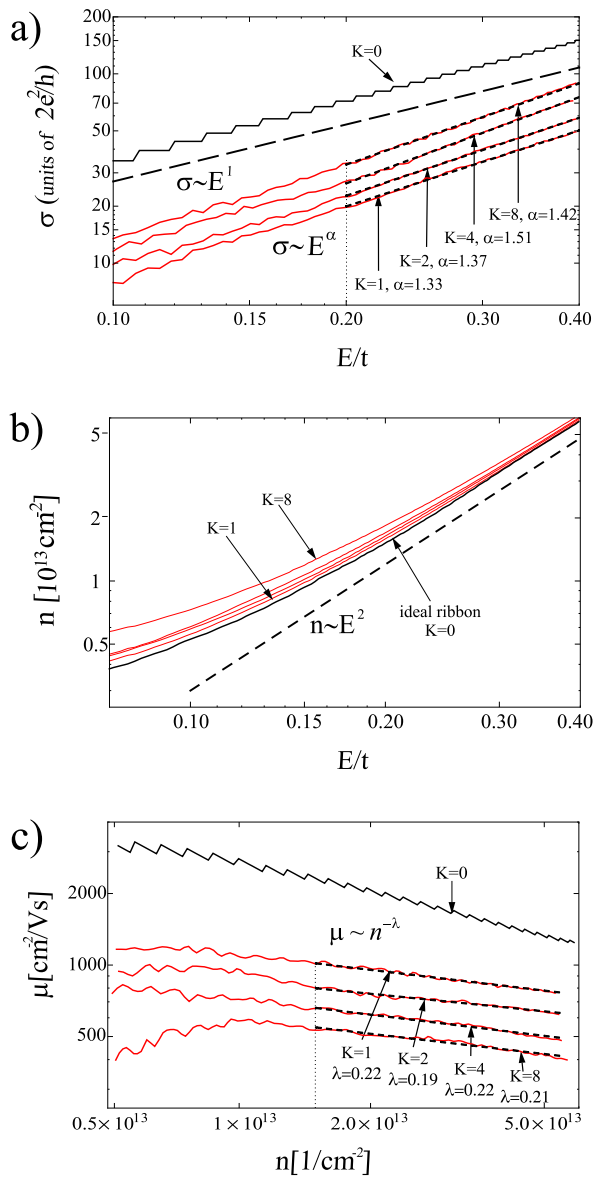}
\caption{(color online) (a) The conductivity and (b) the electron density
\textit{versus} the Fermi energy and (c) the mobility \textit{versus} the
electron density for graphene nanoribbon for different impurity strengths $%
K=1,2,4,8$; $\protect\xi=4a$, $n_{imp}=10^{12}\,$cm$^{-2}$. The ribbon's
dimension is $L\times W= 123 \times 53$ nm$^2 (1000\times 250$ sites). The
dashed lines in (a) and (c) show the fits according to Eqs. (\protect\ref%
{scaling_sigma}) and (\protect\ref{scaling}) for respectively $E>0.2t$ and $%
n\gtrsim 1.5\times 10^{13}$ cm$^{-2}$. (The start of respective fitting
intervals are indicated by vertical dotted lines).}
\label{fig:K}
\end{figure}

For low energies close to the charge neutrality point $E=0$ the DOS of the
system with impurities shows sharps peaks whose positions are strongly
system-dependent, see Fig. \ref{fig:DOS} (a). This is because in this energy
interval the system is in the strong localization regime with the
localization length being smaller than the system size (for an analysis of
this transport regime see Ref. \onlinecite{Xu_bilayer}). In this energy
interval the DOS is strongly system-dependent because it reflects a
particular configuration of the impurity potential. This explains some
differences in the electron densities for different ribbons in Fig. \ref{fig:DOS} (b)
(especially for the shortest one with $L=250$ sites), because the
integration in Eq. (\ref{eq:e02}) includes all the energies (below Fermi energy), including those
close to $E=0$ where the DOS can show strong system-specific
fluctuations.

Having calculated the electron density and the conductivity, we are in a
position to discuss the mobility of the system at hand. The mobility of the
ideal ribbons of different lengths $L$ as a function of the electron density
is shown in Fig. \ref{fig:mobility} (a). This corresponds to the ballistic
regime with $\lambda =0.5$. For a given electron density the mobility of the
ballistic ribbons is proportional to the ribbon length, $\mu \sim L$
(because $\mu =\sigma /en$ and $\sigma \sim L,$ see Fig. \ref%
{fig:conductivity} (c)).

Let us now investigate how the mobility of a ribbon evolves as we go from
the quasi-ballistic to the diffusive regime by increasing the system size.
Figure \ref{fig:mobility} (b) shows the mobility of the ribbon with
impurities as a function of the electron density. As expected, this
dependence satisfies Eq. (\ref{scale_mobility}). The dependence $\lambda
=\lambda (L)$ is shown in the inset to Fig. \ref{fig:mobility} (b). This
dependence clearly demonstrates that we approach the diffusive regime (with
the expected value of $\lambda =0)$ as the length of the ribbons becomes
sufficiently long ($L\gtrsim 740$ nm). This is fully consistent with the
behavior of the conductivity discussed above which exhibits a transition to
the diffusive transport regime as the length of the ribbon increases.

The calculated dependence $\lambda =\lambda (L)$ is also consistent with the
energy dependence of the conductivity which approaches the quadratic
behavior $\sigma \sim E^{2}$ in the diffusive regime (see inset to Fig. \ref%
{fig:conductivity} (b)). Because $n\sim $ $E^{2}$ regardless of the regime
(ballistic, quasi-ballistic, or diffusive), the mobility $\mu =\sigma /en$
becomes independent on the energy (and thus on the electron density
with $\lambda =0$) only when $\sigma \sim E^{2}.$ To illustrate this we in
Fig. \ref{fig:mobility} (c) present the dependence $\sigma =\sigma (n)$
which is plotted by combining previously calculated dependencies $\sigma
=\sigma (E)$ and $n=n(E).$ As expected from Eqs. (\ref{eq:e03}) and (\ref%
{scale_mobility}) it follow a dependence
\begin{equation}
\sigma (n)\sim n^{\gamma },  \label{scale_sigma}
\end{equation}%
where the exponent $\gamma $ ranges from $1$ in the diffusive limit to $0.5$
in the ballistic limit.

All the results for the conductivity, electron density and the mobility in the
graphene ribbons presented above correspond to the case of one
representative impurity strength $n_{imp}=10^{12}$cm$^{-2}$, $K=2$, $\xi =4a$%
. It is important to stress that the scaling laws discussed above are rather
insensitive to a particular realization of the potential configuration or
the impurity strength provided that the system is in the transport regime
when the localization length is larger than the ribbon size. This is
illustrated in Fig. \ref{fig:K} showing the conductivity, the electron density and
the mobility in graphene ribbons for different impurity strength $K=1,2,4,8$.
As expected, the electron density is similar for all ribbons, especially for high energies, 
see Fig. \ref{fig:K} (b). When the impurity strength increases the
conductivity apparently decreases, see Fig. \ref{fig:K} (a). This decrease
of the conductivity leads also to the decrease of the mobility as shown in
Fig. \ref{fig:K} (c). However, the exponents in the scaling laws (\ref%
{scaling_sigma}) and (\ref{scaling}) are not particularly sensitive to the
variation of the impurity strength $K.$ A small difference in scaling
exponents for different impurity strengths has a statistical origin and this
difference diminishes as a number of impurity configurations used in
calculations of each curve is increased.

Let us now use our results to discuss available experimental data. In
experiments, the electron density dependence of the mobility, $\mu =\mu (n)$, and the
conductivity, $\sigma =\sigma (n)$, are accessible \cite{Bolotin,Du}. The
dependencies (\ref{scale_mobility}), (\ref{scale_sigma}) can therefore be
used to extract information about the transport regime for the system at
hand. For example, for the mobility the exponent $\lambda =0.5$ would
correspond to a purely ballistic transport regime, whereas $\lambda =0$
would describe a purely diffusive one. An intermediate exponent $0<\lambda
<0.5$ would indicate the quasi-ballistic transport regime; the more close
the value of $\lambda $ to $0$, the more diffusive the system is. Similar
arguments applies to the conductivity where the corresponding exponent $%
\gamma $ lies between 0.5 and 1. For example, the mobility measured by Du
\textit{et al}. \cite{Du} corresponds to $\lambda =0.5$ suggesting a purely
ballistic transport regime. The electron density dependence of the mobility and
conductivity in graphene ribbons was also studied by Bolotin \textit{et al}.%
\cite{Bolotin} The measured conductivity follows the sublinear behavior (\ref%
{scale_sigma}) with $\gamma <1$. They attributed the deviation from the
linear dependence to the effect of the short-range scattering. Taking into
account that the mean free path in their device is comparable to the device
dimension, we can provide an alternative interpretation of their findings
arguing that the observed in \cite{Bolotin} sublinear behavior represents a
strong evidence of the quasi-ballistic transport regime.

\section{Conclusions}

In the present study we perform numerical calculations of the conductance of
graphene ribbons based on the Landauer formalism and the tight-binding
\textit{p}-orbital Hamiltonian including the effect of warping of graphene
and realistic long-range impurity potential. The effect of warping is
included in our model by modification of the nearest neighbor hopping
integrals resulting from stretching/contraction of the surface and the $\pi
-\sigma $ rehybridisation. We find that the modification of the
next-neighbor hopping due to the warping of the graphene surface has a
negligible effect on the conductance in comparison to the effect charged
impurities even for moderate strength and concentration.

The main focus of our study is a transition from the ballistic to the
diffusive transport regime in realistic graphene ribbons with long-range
impurities which occurs as the size of the system increases. We keep in our
study the width of the ribbon $W$ constant and increase the ribbon length $%
L, $ such that the diffusive regime is achieved when $L/W\gg 1$. We however
expect that the results and conclusions presented in this paper would remain
valid even for bulk diffusive samples with $L/W\sim 1,$ as soon as the
localization length exceeds the size of the system.

We demonstrated that the conductivity of graphene ribbons follows a
power-law dependence $\sigma \sim E^{\alpha }$ with $1\leq \alpha \lesssim
2. $ The case $\alpha =1$ corresponds to the ballistic regime whereas $%
\alpha =2 $ corresponds to the diffusive regime which is reached for
sufficiently long ribbons. In the ballistic regime the conductivity scales
linearly with the length of the system $L$, whereas in the diffusive regime
the conductivity saturates with $L$.

We find that the average electron density in the ribbons with impurities is practically not
 reduced in comparison to the ideal ribbons, and follows the same quadratic
dependence $n\sim E^{2}$ regardless of the transport regime (ballistic,
quasi-ballistic or diffusive). This behavior is consistent with the exponent
$\alpha =2$ reached in the diffusive case, because in this case the mobility
$\mu =\sigma /en$ becomes independent on the energy (and hence on the
electron density) as expected for the diffusive regime.

In experiments the electron density dependence of the mobility, $\mu =\mu (n)$, is
accessible. We find that the mobility of graphene ribbons varies as $\mu
(n)\sim n^{-\lambda },$ with $0\leq \lambda \lesssim 0.5.$ The exponent $%
\lambda $ depends on the size of the system with $\lambda =0.5$
corresponding to short ribbons in the ballistic regime, whereas the
diffusive regime $\lambda =0$ (when the mobility is independent on the
electron density) is reached for sufficiently long ribbons. Our results can be used
for the interpretation of the experimental data when the value of the
parameter $\lambda $ can be used to distinguish the transport regime of the
system (i.e. ballistic, quasi-ballistic or diffusive). The corresponding
electron density dependence for the conductivity is, $\sigma (n)\sim n^{-\gamma },$%
where the exponent $\gamma $ ranges from $1$ in the diffusive limit to $0.5$
in the ballistic limit. Based on our findings we discuss the available
experiments and provide an alternative interpretation of some experimental
conclusions.\cite{Bolotin,Du}

Our calculations also demonstrate that in the quasi-ballistic regime (which
corresponds to many experimental studies) the mobility and the conductivity
of the structure at hand strongly depend on the system size. Therefore in
this regime the conductivity does not exist as a local property and the
mobility can not be considered as a well-defined material parameter because
of its dependence on the system size.

\begin{acknowledgements}
J.W.K. was supported by Polish Ministry of Science and Higher Education within the
program \textit{Support of International Mobility, 2nd edition}.  A.A.S. and I.V.Z.
acknowledge the support from the Swedish Research Council (VR), Swedish Institute (SI)
and from the Swedish Foundation for International Cooperation in Research and Higher
Education (STINT) within the DAAD-STINT collaborative grant. H.X. and T.H. acknowledge
financial support from the German Academic Exchange Service (DAAD) within the DAAD-STINT
collaborative grant.
\end{acknowledgements}

\appendix

\section{Electron conductivity, mobility and electron density in the ballistic regime}

\begin{figure}[tbp]
\centering
\includegraphics[keepaspectratio, width=0.8\columnwidth]{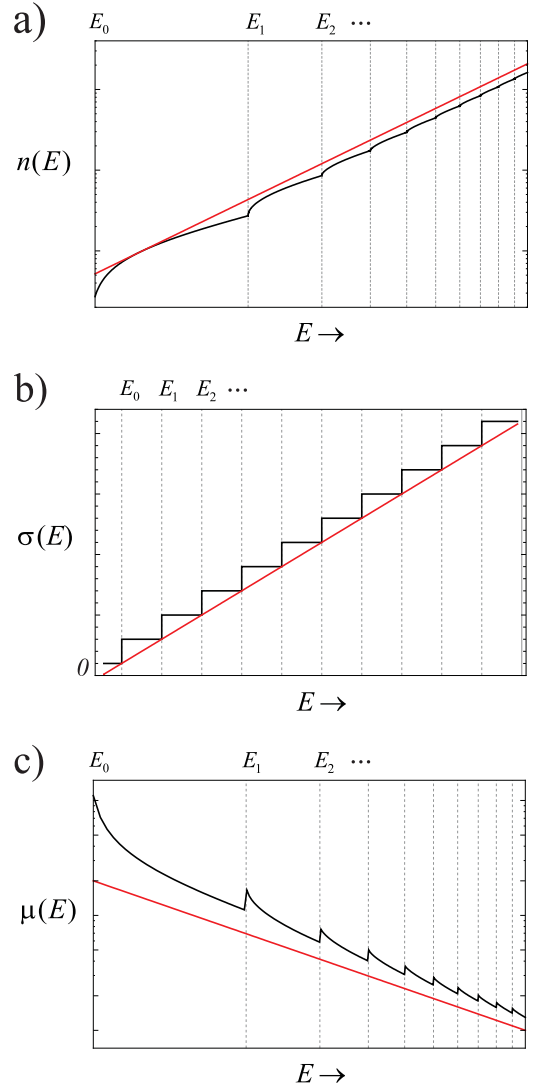}
\caption{ (Color online) (a) The electron density, (b) the
conductivity (c) and the mobility as a functions of the Fermi energy for a
zigzag nanoribbon. Red lines correspond to the approximate expressions (%
\protect\ref{eq:a_n_approx}), (\protect\ref{eq:a_sigma_approx}), (\protect\ref{eq:a_mu}), whereas
the black lines to the exact expressions (\protect\ref{eq:a_n}), (\protect
\ref{eq:a_sigma}). }
\label{fig:appendix}
\end{figure}

In the low-energy limit close to the charge neutrality point $E=0$ the
electron density of the zigzag graphene nanoribbon of the width $N$
reads,\cite{Shylau}
\begin{equation}
n(E)=\frac{4}{\pi \sqrt{3}}\frac{1}{t_{0}a_{0}}\sqrt{E^{2}-E_{m}^{2}}\;\theta (|E|-|E_{m}|),\label{eq:a_n}
\end{equation}
where the threshold energies have the form,
\begin{equation}
E_{m}=\frac{3\pi }{8} t_{0}a_{0}\frac{1}{W}(m+\tfrac{1}{2}), \;m=0,1,2,\ldots .\label{eq:a_ener}
\end{equation}%
Noticing that the function $\sqrt{E^{2}-E_{m}^{2}}$ approaches $E$ for $E\gg
E_{m}$ we can write the electron density in the approximate form:
\begin{equation}
n(E)\approx \frac{4}{\pi \sqrt{3}}\frac{1}{t_{0}a_{0}} E\sum_{m}\;\theta (|E|-|E_{m}|). \label{eq:a_n_approx1}
\end{equation}
Summation of $\theta $-functions in Eq. (\ref{eq:a_n_approx1}) gives a number of
propagating modes at the given energy. Expressing this number with energy by making use of
Eq. (\ref{eq:a_ener}) and approximating $E\approx E_{m}$ we obtain,
\begin{equation}
n(E)\approx \frac{32}{\pi^{2}3\sqrt{3}}\frac{1}{t_{0}^{2}a_{0}^{2}} W E^{2}.  \label{eq:a_n_approx}
\end{equation}%
A comparison of the approximate expression (\ref{eq:a_n_approx}) with the exact one (%
\ref{eq:a_n}) is shown in Fig. \ref{fig:appendix}(a).

In the low-energy limit close to the charge neutrality point $E=0$ the
conductivity of the zigzag graphene nanoribbon of the width $N$ reads,\cite%
{onipko}
\begin{equation}
\sigma(E)=\frac{2e^{2}}{h}\frac{L}{W}\left(\sum_{m}2\theta(|E|-|E_{m}|)-1\right).  \label{eq:a_sigma}
\end{equation}%
Using similar approximations as above we obtain,
\begin{equation}
\sigma(E)\approx \frac{2e^{2}}{h} \frac{16}{3\pi}\frac{1}{ t_{0} a_{0}}L E.
\label{eq:a_sigma_approx}
\end{equation}%
Finally, substituting Eq. (\ref{eq:a_sigma_approx}) and (\ref{eq:a_n_approx}) into the
definition of the mobility we obtain,
\begin{equation}
\mu=\frac{\sigma}{e n}\approx \frac{e}{h}\pi\sqrt{3}t_{0} a_{0} \frac{L}{W}E^{-1}
=\frac{e}{h}\frac{4\sqrt{2}}{\sqrt[4]{3}}\frac{L}{\sqrt{W}}n^{-0.5}.  \label{eq:a_mu}
\end{equation}%
A comparison of the approximate expression for the conductivity and the
mobility with the exact ones is shown in Fig. \ref{fig:appendix} (b), (c).

It should be noted that the expressions (\ref{eq:a_n}), (\ref{eq:a_n_approx1}), (\ref{eq:a_n_approx}) for the
electron density $n$ do not include a contribution from the edge states existing in the
zigzag nanoribbons for the energies close to $E=0$. Because of this the electron density of the
nanoribbon shown in Fig. \ref{fig:appendix} vanishes at $E=0$ and the mobility $\mu $
exhibits a singularity. Accounting for the contribution from the edge state in exact
numerical calculations leads to the finite values of $n$ and $\mu $ at $E=0.$ However,
for high energies sufficiently away the charge neutrality point this contribution does
not practically affect $n$ and $\mu $, which justifies the utilization of expressions
(\ref{eq:a_n})-(\ref{eq:a_mu}).


\end{document}